
\baselineskip=16pt
\vskip 18pt
\noindent
{\bf FINITE DIMENSIONAL REPRESENTATIONS OF THE QUANTUM SUPERALGEBRA
$U_q[gl(3/2)]$ IN A REDUCED $U_q[gl(3/2)] \supset U_q[gl(3/1)] \supset
U_q[gl(3)]$ BASIS}

\leftskip 2cm
\vskip 32pt
\noindent
T. D. Palev\footnote*{Permanent address: Institute for Nuclear Research
and Nuclear Energy, 1784 Sofia, Bulgaria; E-mail
palev@bgearn.bitnet}

\noindent
Applied Mathematics and Computer Science, University of Ghent,
B-9000 Gent, Belgium

\noindent
and

\noindent
International Centre for Theoretical Physics, 34100 Trieste, Italy

\vskip 12pt
\noindent
N. I. Stoilova*

\noindent
Applied Mathematics and Computer Science, University of Ghent,
B-9000 Gent, Belgium

\vskip 2cm

{\bf Abstract.} For generic $q$ we give expressions for the
transformations of all essentially typical finite-dimensional
modules of the Hopf superalgebra $U_q[gl(3/2)]$. The latter is a
deformation of the universal enveloping algebra of the Lie
superalgebra $gl(3/2)$. The basis within each module is similar to the
Gel'fand-Zetlin basis for $gl(5)$. We write down expressions for
the transformations of the basis under the action of the
Chevalley  generators.

\vskip 18mm
\noindent
PACS numbers: 02.20, 03.65, 11.30

\vskip 25mm
\leftskip 12pt
\noindent
This letter is devoted to the  study of a subclass of representations
of the quantized associative superalgebra $U_q[gl(3/2)]$, which
we call essentially typical (see Definition 1). The motivation
to consider this particular algebra stems, firstly,
from the observation
that it is relatively simple and nevertheless its representation
theory remains so far undeveloped explicitly and, secondly,
that it carries  all main
features of the general $U_q[gl(n/m)]$ superalgebra.
The hope is, therefore, that  $U_q[gl(3/2)]$
could serve as a generic example to tackle as  a next step
the representation theory of the deformed superalgebra $gl(n/m)$.

To begin with we recall that a quantum algebra
$U_q[G]$ associated with the Lie (super)algebra $G$ is a
deformation of the universal enveloping algebra $U[G]$ of $G$,
preserving its Hopf algebra structure. The first
example of this kind was found for $G=sl(2)$ [1] endowed with
a Hopf structure given by
Sklyanin [2]. An example of a quantum superalgebra, namely
the LS $osp(1/2)$, was first considered by Kulish [3] and it
was soon generalized to an arbitrary Kac-Moody superalgebra with
a symmetrizable generalized Cartan matrix [4]; an
independent approach for basic Lie superalgebras was given
in [5-7].

Beginning with $U_q[osp(1/2)]$ [8] the representation theory
of the quantum superalgebras became also a field of growing interest.
Various oscillator representations of all infinite
series of the basic Lie superalgebras have been found (see,
for instance, Refs. 5, 7, 9, 10).  Nevertheless the results in
this respect are still modest, the main reason being that even
in the nondeformed case the representation theory of the best
known, namely the basic Lie superalgebras (LS's) is far from being
complete. Even the problem of obtaining
character formulae for all atypical modules is
still not completely solved (see [11] and the references therein).
Apart from the lowest rank LS's, explicit expressions for the
transformations of all finite-dimensional modules have been
given so far only for the class  $gl(n/1), \quad n=1,2,\ldots $ [12].
By explicit we understand
what is usually meant in physics, namely introducing a basis in
the representation space and writing down explicit expressions
for the transformations of the basis under the action of the
generators ( or, which is the same, giving expressions for
the matrix elements of the generators with respect to the
selected basis). The results from [12] have been used in [13] in order
to show that each finite-dimensional irreducible $gl(n/1)$
module can be deformed to an irreducible $U_q[gl(n/1)]$ module.
In a close  analogy with the results of Jimbo [14]
for $U_q[A_n]$, expressions were given for the transformations of
the deformed $U_q[gl(n/1)]$ modules in terms of the undeformed
Gel'fand-Zetlin basis. The present investigation is much along
the same line. We deal however only with the representations we
know explicitly, namely the essentially typical representations
of $gl(3/2)$.

Before going to the deformed case, we recall that the Lie
superalgebra (LS) $gl(3/2)$ is an extension of the basic LS
$sl(3/2)$ [15] by a one dimensional center with  generator
$I$. A convenient basis in $gl(3/2)$ is given by the Weyl
generators $e_{ij}, \, i,j=1,2,3,4,5$, which satisfy the relations

$$[e_{ij},e_{kl}]=\delta_{jk}e_{il}
-(-1)^{\theta_{ij}\theta_{kl}}\delta_{il}e_{kj}. \eqno (1) $$

\noindent
Here and everywhere in the text

$$[e_{ij},e_{kl}] \equiv e_{ij}e_{kl}
-(-1)^{\theta_{ij}\theta_{kl}}e_{kl}e_{ij} , \eqno (2)  $$
and
$$\theta_{ij}=\theta_{i}+\theta_{j}, \quad
\theta_{i}=0 \;for \; i=1,2,3, \quad \theta_{i}=1
\; for\; i=3,4. \eqno (3)  $$

\noindent
The  ${\bf Z}_2$-grading on $gl(3/2)$ is imposed from the
requirement that $e_{ij}$ is an even (respectively an odd)
generator, if $\theta_{ij}$ is an even (respectively odd) number.
The central element $I$ reads

$$I=e_{11}+e_{22}+e_{33}+e_{44}+e_{55} \eqno(4) $$

\noindent
and $sl(3/2)$ is the factor algebra of $gl(3/2)$ with respect to
the ideal generated by $I$. In the lowest $ 5\times 5$
representation $e_{ij}$ is a matrix with 1 on the cross of the
$i^{th}$ row and the $j^{th}$ column and zeros elsewhere and $I$
is the unit matrix.

The universal enveloping algebra $U[gl(3/2)]$
of $gl(3/2)$ is the free associative algebra of all generators
$e_{ij}$ with relations (1). A representation of the LS
$gl(3/2)$ is by definition a representation of $U[gl(3/2)]$
viewed as a ${\bf Z}_2 \; (=\{0, 1\})$-graded associative
algebra, i.e., as  associative superalgebra.

The irreducible finite-dimensional
modules (=representation spaces) of any basic LS are either
typical or atypical [16]. It is convenient to consider any
irreducible $sl(3/2)$ module $W$ as an irreducible $gl(3/2)$ module,
setting  $I$ proportional to the identity operator in
$W$.  In this sense one can speak about typical or
atypical $gl(3/2)$ modules. More precisely, by a typical
(resp. atypical) $gl(3/2)$ module we understand a $gl(3/2)$ module $W$,
which remains typical (resp. atypical) with respect to $sl(3/2)$ and
for which $I$ is proportional to the identity operator in $W$.
A subclass of the typical modules will be relevant for us.

{\it Definition 1}[17]. A typical $gl(3/2)$ module $W$ is said to
be essentially typical, if it decomposes into a direct sum of only
typical $gl(3/1)$ modules.

In the present letter we show that for generic values of $q$
(=$q$ is not root of unity) any essentially typical $gl(3/2)$ module
$W$ can be deformed to an irreducible $U_q[gl(3/2)]$ module, which
we call also essentially typical. We proceed to describe in some more
detail a basis  within each typical module (and, more generally,
- within a larger class of finite-dimensional induced modules, namely
the  Kac modules [18]), since the same basis
will be used   also in the deformed case.

Let $e_{11}, e_{22}, e_{33}, e_{44}, e_{55}$ be a basis in the
Cartan subalgebra $H$ and $e^1,e^2, e^3, e^4, e^5 $ be
the basis dual to it in  $H^*$. Denote by
$W([m]_5)$ the Kac module with  highest weight

$$[m]_5 =m_{15}e^1+m_{25}e^2+m_{35}e^3+m_{45}e^4+
m_{55}e^5 \equiv
[m_{15},m_{25},m_{35},m_{45},m_{55}], \eqno (5) $$

\noindent
and let
$$l_{i5}=m_{i5}-i+4 \;for \;i=1, 2, 3 \quad and \quad
l_{i5}=-m_{i5}+i-3 \;for \;i=4,5. \eqno (6)  $$

{\it Proposition 1} [17]. The Kac modules $W([m]_5)$ are in one to
one correspondence with the set of all complex coordinates of the
highest weight, which satisfy the conditions:

$$m_{15}-m_{25}, m_{25}-m_{35},
m_{45}-m_{55}\in {\bf Z}_+\, {\rm
[= \;all  \; nonnegative \; integers]}. \eqno(7)  $$

\noindent
The module $W([m]_5)$ is typical if and only if all
$l_{15},l_{25},l_{35},l_{45},l_{55}$ are different.
The module $W([m]_5)$ is essentially typical if and only if
$l_{15},l_{25},l_{35}\neq  l_{45}, l_{45}+1,l_{45}+2,
\ldots ,l_{55}$.

Let $(m)$ be a pattern consisting of 15 complex numbers $m_{ij}, \;
i\leq j=1,2,3,4,5$, ordered as in the usual Gel'fan-Zetlin basis
for $gl(5)$ [19], i. e.,

$$(m)\equiv \left(\matrix
{m_{15}& \,m_{25}& \,m_{35}& \,m_{45}& \,m_{55}\cr
 m_{14}& \,m_{24}& \,m_{34}& \,m_{44}& \cr
 m_{13}& \,m_{23}& \,m_{33}& & \cr
 m_{12}& \,m_{22}& & & \cr
 m_{11}& & & & \cr }\right). \eqno(8)  $$

{\it Proposition 2} [20]. The set of all patterns (8),
whose entries satisfy the conditions

(1) $m_{15},m_{25},m_{35},m_{45},m_{55}$ are fixed and the same
for all patterns,

(2) For each $i=1,2,3$ and $p=4, 5 \quad
    m_{ip}-m_{i,p-1} \equiv \theta_{i,p-1} \in {\bf Z}_2$,

(3) For each $i=1,2$ and $p=4, 5
    \quad m_{ip}-m_{i+1,p} \in {\bf Z}_+$, \hskip 193pt (9)

(4) For each $i\leq j=1,2 \quad m_{i,j+1}-m_{ij},\,
    m_{ij}-m_{i+1,j+1} \in {\bf Z}_+ $,

(5) $m_{45}-m_{44},\;m_{44}-m_{55} \in {\bf Z}_+   $,

\noindent
constitute a basis in the Kac module $W([m]_5)$, which we refer
to as a GZ-basis.

The quantum superalgebra $U_q[gl(3/2)]$, we consider here, is
a one parameter Hopf deformation of the universal enveloping
algebra of $gl(3/2)$
with a deformation parameter $q=e^{\hbar}$. As usually, the limit
$q \rightarrow 1 $ {}( $\hbar \rightarrow 0$) corresponds to the
nondeformed case. More precisely, $U_q[gl(3/2)] \equiv U_q$ is a free
associative algebra with unity , generated by $e_i, \; f_i$ and
$k_j \equiv q^{h_j/2},\quad i=1,2,3,4, \quad j=1,2,3,4,5 $, which
satisfy (unless otherwise stated the indices below run over all
possible values)

\noindent
1. The Cartan relations

$$k_ik_j=k_jk_i,\quad \quad k_ik_i^{-1}=k_i^{-1}k_i=1; \eqno (10) $$

$$k_ie_jk_i^{-1}=q^{{1 \over 2}(\delta_{ij}-\delta_{i,j+1})}e_j,
\quad
k_if_jk_i^{-1}=q^{-{1 \over 2}(\delta_{ij}-\delta_{i,j+1})}f_j;
\eqno(11) $$

$$e_if_j-f_je_i=0 \; i\neq j; \quad \quad
e_if_i-f_ie_i=(q-q^{-1})^{-1}(k_i^2k_{i+1}^{-2}-
k_{i+1}^2k_i^{-2}),\; i=1,2,4; \eqno(12)   $$

$$e_3f_3+f_3e_3=(q-q^{-1})^{-1}(k_3^2k_4^2-k_3^{-2}k_4^{-2});
\eqno(13)  $$

\noindent
2. The Serre relations for the positive simple root vectors
(E-Serre relations)

$$e_ie_j=e_je_i, \;  \vert i-j \vert \neq 1;\quad
e_3^2=0;   \eqno(14) $$

$$e_i^2e_{i+1}-(q+q^{-1})e_ie_{i+1}e_i+e_{i+1}e_i^2=0,
\quad i=1,2; \eqno(15)  $$

$$e_{i+1}^2e_i-(q+q^{-1})e_{i+1}e_ie_{i+1}+e_ie_{i+1}^2=0
\quad i=1,3; \eqno(16)   $$

$$e_3e_2e_3e_4+e_2e_3e_4e_3+e_3e_4e_3e_2+e_4e_3e_2e_3
-(q+q^{-1})e_3e_2e_4e_3=0; \eqno(17)  $$

\noindent
3. The relations obtained from eqs.(14-17) by replacing
everywhere  $e_i$ by $f_i$ (F-Serre relations).

\vskip 12pt
\noindent
The eq.(17) is the extra Serre
relation, discovered recently [21-23] and reported  by
Tolstoy, Scheunert and Vinet on the II International Wigner
Symposium (July 1991, Goslar, Germany).

The ${\bf Z}_2$-grading on $U_q[gl(3/2)]$ is defined from the
requirement that the only odd generators are $e_3$ and $f_3$.
$U_q$ is a Hopf algebra with a counity $\varepsilon$,
a comultiplication $\Delta$ and an antipode $S$ defined as

$$\varepsilon(e_i)=\varepsilon(f_i)=\varepsilon(k_i)=0,\hskip 237pt
\eqno(18)   $$

$$\vcenter{\halign{ $#$ \hfil \cr
\Delta(k_i)=k_i \otimes k_i, \cr
\Delta(e_i)=e_i \otimes k_ik_{i+1}^{-1}+
k_i^{-1}k_{i+1} \otimes e_i,\; i\neq 3, \quad
\Delta(e_3)=e_3 \otimes k_3k_4+
k_3^{-1}k_4^{-1} \otimes e_3, \cr
\Delta(f_i)=f_i \otimes k_ik_{i+1}^{-1}+
k_i^{-1}k_{i+1} \otimes f_i,\; i\neq 3, \quad
\Delta(f_3)=f_3 \otimes k_3k_4+
k_3^{-1}k_4^{-1} \otimes f_3, \cr
}} \eqno(19)$$

$$\hskip 2pt S(k_i)=k_i^{-1}, \; S(e_j)=-qe_j, \;
S(f_j)=-q^{-1}f_j,  j\neq 3,\; S(e_3)=-e_3,\;S(f_3)=-f_3. \eqno(20) $$

Now we are ready to state our main result. To this end denote by
$(m)_{\pm ij}$ a pattern obtained from the GZ pattern $(m)$
after a replacement of $m_{ij}$ with $m_{ij} \pm 1$  and let

$$l_{ij}=m_{ij}-i+4 \;for \;i=1, 2, 3 \quad and \quad
l_{ij}=-m_{ij}+i-3 \;for \;i=4,5, \quad
[x]={{q^{x}-q^{-x}}\over {q-q^{-1}}}. \eqno (21)  $$

{\it Proposition 3.} For generic values of $q$
every essentially typical
$gl(3/2)$ module $W([m]_5)$ can be deformed to an irreducible
$U_q[gl(3/2)]$ module. The transformations of the basis under
the action of the algebra generators read
$$\displaylines{\quad
k_i(m)=q^{{1\over 2}(\sum_{j=1}^i m_{ji}-
\sum_{j=1}^{i-1} m_{j,i-1}) }(m), \quad \quad i=1,2,3,4,5,
\hfill (22) \cr
}$$

$$\displaylines{\quad
e_k(m)=\sum_{j=1}^k \left(-
{\prod_{i=1}^{k+1} [l_{i,k+1}-l_{jk}]
\prod_{i=1}^{k-1} [l_{i,k-1}-l_{jk}-1]
\over \prod_{i\neq j=1}^k [l_{ik}-l_{jk}]
[l_{ik}-l_{jk}-1] } \right)^{1/2}(m)_{jk},\quad k=1,2,
\hfill (23) \cr
}$$

$$\displaylines{\quad
f_k(m)=\sum_{j=1}^k \left(-
{\prod_{i=1}^{k+1} [  l_{i,k+1}-l_{jk}+1  ]
\prod_{i=1}^{k-1} [  l_{i,k-1}-l_{jk}  ]
\over \prod_{i\neq j=1}^k [  l_{ik}-l_{jk}+1  ]
[  l_{ik}-l_{jk}  ] } \right)^{1/2}(m)_{-jk},\quad k=1,2,
\hfill (24) \cr
}$$

$$\displaylines{\quad
e_3(m)=\sum_{i=1}^3 \theta_{i3}(-1)^{i-1}
(-1)^{\theta_{13}+ \ldots +\theta_{i-1,3} }
\left(\prod_{k=1}^2 [  l_{k2}-l_{i4} ]
\over \prod_{k\neq i=1}^3 [  l_{k4}-l_{i4} ]
\right)^{1/2} (m)_{i3},
\hfill (25) \cr
}$$
$$\displaylines{\quad
f_3(m)=\sum_{i=1}^3 (1-\theta_{i3})(-1)^{i-1}
(-1)^{\theta_{13}+ \ldots +\theta_{i-1,3} }[l_{i4}-l_{44}]
\left(\prod_{k=1}^2 [  l_{k2}-l_{i4} ]
\over \prod_{k\neq i=1}^3 [  l_{k4}-l_{i4} ]
\right)^{1/2} (m)_{-i3},
\hfill (26) \cr
}$$

$$\displaylines{\quad
e_4(m)=\sum_{i=1}^3 \theta_{i4}(1-\theta_{i3})
(-1)^{\theta_{14}+ \ldots +\theta_{i-1,4}+
\theta_{i+1,3}+ \ldots +\theta_{33} }
\prod_{k\neq i=1}^3 \left(
[l_{i5}-l_{k4}]  [l_{i5}-l_{k4}-1] \over
[l_{i5}-l_{k5}][l_{i5}-l_{k3}-1] \right)^{1/2}(m)_{i4}
\hfill \cr
\hskip 53pt +([l_{44}-l_{45}][l_{55}-l_{44}])^{1/2}
\prod_{k=1}^3 {[l_{k4}-l_{44}] [l_{k4}-l_{44}+1] \over
[l_{k5}-l_{44}][l_{k3}-l_{44}+1]} (m)_{44},
\hfill (27) \cr
}$$

$$\displaylines{\quad
f_4(m)=([l_{44}-l_{45}+1][l_{55}-l_{44}-1])^{1/2}(m)_{-44}+
\sum_{i=1}^3 \theta_{i3}(1-\theta_{i4})
(-1)^{\theta_{14}+ \ldots +\theta_{i-1,4}+
\theta_{i+1,3}+ \ldots +\theta_{33} }
\hfill  \cr
\hskip 53pt   \times
{[l_{i5}-l_{45}] [l_{i5}-l_{55}] \over
[l_{i5}-l_{44}-1][l_{i5}-l_{44}]}
\prod_{k\neq i=1}^3 \left(
[l_{i5}-l_{k4}]  [l_{i5}-l_{k4}-1] \over
[l_{i5}-l_{k5}][l_{i5}-l_{k3}-1] \right)^{1/2}(m)_{-i4}.
\hfill (28) \cr
}$$

The proof is straightforward but lengthy. In order to perform it
one has to check that the Cartan relations and the E- and F-Serre
relations hold as operator equations in $W([m]_5)$.
The irreducibility follows from the results of Zhang [24]. It is not
difficult to check it directly and follows from the observation
that for generic $q$ a deformed matrix element in the GZ basis
is zero only if the corresponding nondeformed matrix element
vanishes [see (22)-(28)].

{\it Remark.} If a vector from the rhs of eqs. (22)-(28) does
not belong to the  module under consideration, than
the corresponding term is zero even if the
coefficient in front of it is undefined. If an equal number of
factors in a numerator and a denominator are simultaneously
equal to zero, then one has to cancel them out.

The operators $e_{i-1}, \; f_{i-1}, \;k_{i-1}, \; k_i, \quad
i=1,2, \ldots,n$ generate a chain of Hopf subalgebras
$U_q[gl(1)]$ for $n=1$, $\; U_q[gl(2)]$ for $n=2$,
$\; U_q[gl(3)]$ for $n=3$, $\;U_q[gl(3/1)]$ for $n=4$,
$\; U_q[gl(3/2)]$ for $n=5$,

$$U_q[gl(1)] \subset U_q[gl(2)] \subset U_q[gl(3)]
\subset U_q[gl(3/1)] \subset U_q[gl(3/2)]. \eqno(29)$$

\noindent
The row $[m]_n \equiv (m_{1n},\ldots,m_{1n})$ in a GZ pattern
$(m)$ indicates that $(m)$ belongs to an irreducible  submodule
$W([m]_n) \subset W([m]_5)$ of the subalgebra with the corresponding
$n$. Therefore the decomposition of any essentially typical
module $W([m]_5)$ along the chain (29) is evident and the GZ
basis is reduced with respect to this chain.
More precisely  , each basis
vector $(m)$ is an element from one and only one chain of submodules,

$$(m) \in W([m]_1) \subset W([m]_2) \subset W([m]_3) \subset
W([m]_4) \subset W([m]_5). \eqno(30) $$

Unfortunately we do not know at present how to extend the above
results beyond the class of the essentially typical
representations, although it is known that each irreducible
finite-dimensional representation  of $gl(3/2)$ can be deformed
to a $U_q[gl(3/2)]$ representation [24]. We hope, however,
that the present approach could at least be generalized for
the essentially typical representations of $U_q[gl(n/m)]$,
although technically this could be quite difficult.
It remains still an open question whether one can deform the Kac
modules even for generic $q$.

\vskip 12pt
We are grateful to
Prof. Vanden Berghe for the possibility to work at the
Department of Applied Mathematics and Computer Science,
University of Ghent.
It is a pleasure to thank Dr. J. Van der Jeugt
for stimulating discussions and  introducing us to  Maxima,
which was  of great help to us.
One of us (T. D. P.) is thankful to Prof. Abdus Salam for the kind
hospitality at the International Center for Theoretical Physics,
Trieste.

The work was supported by the European Community, contract
No ERB-CIPA-CT-92-2011 (Cooperation in Science and Technology
with Central and Eastern European Countries) and
grant $\Phi - 215$ of the Committee of Science in Bulgaria.

\vskip 20mm

\noindent
{\bf References}

\vskip 12pt

\settabs\+[11] & I. Patera, T. D. Palev, Theoretical interpretation of the
   experiments on the elastic \cr

\+[1] & Kulish P P and Reshetikhin N Yu 1980 {\it Zapiski nauch.
        semin. LOMI} {\bf 101} 112 \cr

\+[2] & Sklyanin E K 1985 {\it Usp. Math. Nauk} {\bf 40} 214 \cr

\+[3] & Kulish P P 1988 {\it Zapiski nauch. semin. LOMI} {\bf
        169} 95 \cr

\+[4] & Tolstoy V N 1990 {\it Lect. Notes in Physics} {\bf 370} 118
        Berlin, Heidelberg, New York: Springer \cr

\+[5] & Chaichian M and Kulish P P 1990 {\it Phys. Lett B} {\bf
        234} 72 \cr

\+[6] & Bracken A J, Gould M D and Zhang R B 1990 {\it Mod. Phys. Lett.
        A} {\bf 5} 831 \cr

\+[7] & Floreanini R, Spiridonov V and Vinet L 1991
        {\it Commun Math. Phys.} {\bf 137} 149 \cr

\+[8] & Kulish P P and Reshetikhin N Yu 1989
        {\it Lett. Math. Phys} {\bf 18} 143 \cr

\+[9] & Chaichian M, Kulish P P  and Lukierski J 1990 {\it
        Phys. Lett B} {\bf 237} 401 \cr

\+[10]& Floreanini R, Spiridonov V and Vinet L {\it Phys. Lett
        B} {\bf 242} 383 \cr

\+[11]& Van der Jeugt J, Hughes J W B, King R C and Thierry-Mieg
        J 1990 {\it Journ. Math. Phys.} {\bf 31} 2278  \cr

\+    & Hughes J W B, King R C and Van der Jeugt J

        1992 {\it Journ. Math. Phys.} {\bf 33} 470  \cr

\+[12]& Palev T D 1987 {\it Funkt. Anal. Prilozh.} {\bf 21} No3
        85 (in Russian); \cr

\+    & {\it Funct. Anal. Appl.} {\bf 21} 245 (English translation) \cr

\+    & Palev T D 1989 {\it Journ. Math. Phys.} {\bf 30} 1433 \cr

\+[13]& Palev T D and Tolstoy V. N. 1991
        {\it Commun Math. Phys.} {\bf 141} 549 \cr

\+[14]& Jimbo M 1986 {\it Lect. Notes in Physics} {\bf 246} 334
        Berlin, Heidelberg, New York: Springer \cr
\+[15]& Kac V G 1977 {\it Adv. Math.} {\bf 26} 8 \cr

\+[16]& Kac V G 1979 {\it Lect. Notes in Mathematics} {\bf 626} 597.
        Berlin, Heidelberg, New York: Springer \cr

\+[17]& Palev T D 1989 {\it Funkt. Anal. Prilozh.} {\bf 23} No2
        69 (in Russian); \cr
\+    & {\it Funct. Anal. Appl.} {\bf 23} 141 (English translation) \cr

\+[18]& Van der Jeugt J, Hughes J W B, King R C and Thierry-Mieg
        J  1990 {\it Commun. Algebra} {\bf 18} 3453  \cr

\+[19]& Gel'fand I M and Zetlin M L 1950 {\it DAN SSSR} {\bf 71} 825-838
        (in Russian) \cr

\+[20]& Palev T D (unpublished). Some of the results have been
        announced in [17] \cr

\+[21]& Khoroshkin S M and Tolstoy V N 1991
        {\it Commun Math. Phys.} {\bf 141} 599 \cr

\+[22]& Scheunert 1992 {\it Lett. Math. Phys} {\bf 24} 173 \cr

\+[23]& Floreanini R, Leites D A and Vinet L 1991
        {\it Lett. Math. Phys} {\bf 23} 127 \cr

\+[24]& Zhang R B 1993 {\it Journ. Math. Phys.} {\bf 34} 1236 \cr

\end